\documentclass[aps,onecolumn,a4paper,oneside,preprint,tightenlines,eqsecnum,nofootinbib,13 pt]{revtex4}

\usepackage{amsmath,amssymb,amsthm,epsf,epsfig}
\usepackage{amsmath,amssymb}
\usepackage{multirow}
\begin{document}
\title{A note on the $\mathcal{PT}$ invariant periodic potential $V(x)=4 cos^2 x + 4 i V_0 sin 2x$}

\author{Bikashkali Midya}
\thanks{E-mail: bikash.midya@gmail.com~~(B Midya)}
\author{Barnana Roy}
\thanks{E-mail: barnana@isical.ac.in~~(B Roy)}
\author{Rajkumar
Roychoudhury} \thanks{E-mail: raj@isical.ac.in~~( R Roychoudhury)}
\affiliation{
 Physics and Applied
Mathematics Unit, Indian Statistical Institute,\\ Kolkata
700108,~India.}

 \begin{abstract} It is shown that the ${\cal{PT}}$ symmetric Hamiltonian with the periodic potential $V(x) = 4 cos^2 x + 4 i V_0 sin 2x$ can be mapped into a Hermitian Hamiltonian for $V_0<.5$, by a similarity transformation. It is also shown that there exist a second critical point of the potential $V(x)$, apart from the known critical point $V_0=0.5$, for $V_0^c\sim .888437$ after which no part
of the eigenvalues and the band structure remains
real. Relevant physical consequence of this finding has been
pointed out.
\end{abstract}

\maketitle

 Quite recently, the prospect of realizing
complex $\cal {PT}$ symmetric potentials within
 the framework of optics has been suggested \cite{r1,r2,r3,r4,r5,r6,r7,r20,r28,r26}.
 What makes this possible is the formal equivalence
 between the quantum mechanical Schr\"odinger equation and the paraxial equation of diffraction.
In this optical analogy the complex refractive index distribution
plays the role of the optical potential.
 The parity-time condition implies that the real index profile should be even in the transverse direction
 while the loss or gain distribution must be odd. Thus complex optical $\cal {PT}$ potentials can be realized
 by judiciously integrating an asymmetric gain or loss profile on an even index distribution \cite{r1}.
 In these studies it was shown that optical $\cal {PT}$ symmetric materials can lead to behavior
 that is impossible in standard systems. Such effects include double refraction, power oscillation, eigenfunction
 unfolding, non-reciprocal diffraction patterns, bloch oscillations \cite{r1,r2,r4} to mention a few. Recently Guo et al \cite{r3} have demonstrated experimentally passive $\cal {PT}$ symmetry
 breaking within the realm of optics. In ref.\cite{r26}, the first observation of the behavior of a ${\cal{PT}}$ optical coupled system that involves a complex index potential is reported.
  Energy band formation \cite{r9,r21,r22,r27} and spontaneous symmetry breaking thresholds have been investigated in a few $\cal {PT}$ symmetric models even in presence of nonlinearity \cite{r1} and lattice disorder \cite{r20}. In particular, while discussing energy bands due to a complex periodic potential, the authors of ref.\cite{r9} argued from the graphs of $\Delta(E),~E\in \mathbb{R}$ corresponding to these potentials, that some energy bands would appear and disappear under perturbations. Later it was shown that \cite{r27} the appearance and disappearance of such energy bands imply existence of non real spectra due  to the complex deformation of real intervals.

In this short note we shall concentrate on the complex periodic
potential
\begin{equation}
V(x)= 4 cos^2 x+ 4iV_0 sin 2x\label{e1}
\end{equation}
where $V(x+\pi)= V(x)$. This potential is invariant under the dual
action of the space reflection operator $\cal{P}$ and the time
reversal operator $\cal{T}$ \cite{r11}. In general the action of
$\cal{P}$ is defined by the relation $\hat{p}\rightarrow -
\hat{p}$, $\hat{x}\rightarrow -x$ ( $\hat{ p}, \hat{x}$ stand for
momentum and position operators respectively) whereas that of the
time operator $\cal{T}$ by $\hat{p}\rightarrow - \hat{p},
\hat{x}\rightarrow \hat{x}, i \rightarrow -i$ (for more details on
$\cal {PT}$ symmetry see \cite{r12} and references there). The
reason for choosing this particular potential (\ref{e1}) lies in
the fact many physically interesting results \cite{r1,r2} have
been obtained while considering this potential in optical lattice.
In particular, in ref \cite{r2}, by using spectral techniques and detailed numerical calculations the
$\cal {PT}$ threshold $V_0^{th} = 0.5$ has been identified for
this potential. It is found that below this threshold all the
eigenvalues for every band and every Bloch wave number
are real and also all the forbidden gaps are open whereas at the
threshold $V_0^{th} = 0.5$, the spectrum is whole real line and there are no band gaps \cite{r25}. On the other hand, when $V_0
> 0.5$ the first two bands (starting from the lowest band) start to merge together and in doing so they form oval-like structure with a related complex spectrum.
The real as well as the imaginary parts of such a double valued
band when $V_0 = 0.75$ are depicted in Figure 2. We shall argue that one can anticipate by analytical argument that when $V_0<.5$ the ${\cal{PT}}$ symmetry of the potential is unbroken. We shall also see
that there is another critical point $(V_0^c)$ after which
no part of the eigenvalues and the band structure
remains real.

Let us consider the following linear eigenvalue problem frequently
used to study the linear properties of a periodic potential in
optical lattice \cite{r2}
\begin{equation}
H\psi(x) = \frac{d^2\psi(x)}{dx^2}+V(x)\psi(x)=\beta\psi(x)\label{e4}
\end{equation}
where $\beta$ represents the propagation constant (energy eigenvalue) in the periodic
structure and $V(x)$ is given in (\ref{e1}) can be written as:
\begin{equation}
V(x)=\left\{\begin{array}{lll}\displaystyle 2+2\sqrt{1-4V_0^2}~
cos\left(2x-i~ tanh^{-1}2V_0\right),~~~~~~~~V_0< .5\\
\displaystyle 2+2e^{2 i x},~~~~~~~~~~~~~~~~~~~~~~~~~~~~~~~~~~~~~~~~~~~~~~~~~V_0=.5\\
\displaystyle 2+2i \sqrt{4V_0^2-1}~ sin\left(2 x-i~
 tanh^{-1}\frac{1}{2V_0}\right),~~~~~~V_0> .5
\end{array}\right.\label{e2}
\end{equation}

  Recently in ref \cite{r18,r30}, it has been shown that the concept of ${\cal{PT}}$ symmetry has its roots in the theory of pseudo-Hermitian operators . The reality of the spectrum is ensured if the Hamiltonian $H$ is Hermitian with respect to a positive definite inner product $\langle\langle\psi_1|\psi_2\rangle\rangle_{\eta} := \langle\psi_1|\eta|\psi_2\rangle$, $\forall |\psi_1\rangle, |\psi_2\rangle \in {\cal{H}}$ [a Hilbert space in which $H$ is acting], where the Hermitian linear automorphism $\eta: {\cal{H}}\rightarrow {\cal{H}}$ is positive definite. Consequently, $\eta$ pseudo-Hermiticity of the Hamiltonian $H$ is given by
\begin{equation}
H^\dagger = \eta H \eta^{-1}\label{e11}
\end{equation}

 It has also been shown that \cite{r40} for any ${\cal{PT}}$ symmetric Hamiltonian $H$ having real eigenvalues there exist an equivalent Hermitian Hamiltonian $h$ by means of a similarity transformation $h = \rho H \rho^{-1} $, where $\rho = \sqrt{\eta}$ is unitary [in the sense that it is Hermitian and for any pair of state vectors $|\psi\rangle, |\phi\rangle$ : $\langle\langle\psi|\phi\rangle\rangle_{\eta} = \langle\psi|\eta \phi\rangle = \langle\psi|\rho^2 \phi\rangle =\langle\rho\psi|\rho \phi\rangle$ holds] which maps ${\cal{H}}_{phys}\rightarrow {\cal{H}}$ [${\cal{H}}_{phys}$ being the Hilbert space equipped with the inner product defined above]. The Hermitian Hamiltonian $h$ is equivalent to $H$ and that they have same eigenvalues.\\

\vspace{2 cm}
Now it is to be observed from (\ref{e4}) and (\ref{e2}) that \\
 (i) for $V_0< 0.5$ the Hamiltonian defined in (\ref{e4}) satisfy the
relations{\footnote{In ref.\cite{r8}, it has been shown that the Hermitian linear automorphism $\eta=e^{-\theta p}$ affects an imaginary shift of the co-ordinate: $e^{-\theta p} x e^{\theta p} = x + i \theta.$}} $H^\dagger = \eta H \eta^{-1}$ and $h = \rho H \rho^{-1}$, where $h\equiv\frac{d^2}{dx^2}+2+2\sqrt{1-4V_0^2}~ cos~ 2x$,~ $\rho= e^{-\frac{\theta}{2} p}$ [is a unitary operator in the sense stated earlier], $\eta= \rho^2 = e^{-\theta p}$ [is a positive definite operator\footnote{$\eta$ is square of a Hermitian operator, so it is positive.}], $\theta= tanh^{-1} 2V_0 \in \mathbb{R}$ and $p\equiv
-i\frac{d}{dx}.$

Hence the Hamiltonian $H$ is $\eta$-pseudo-Hermitian with respect to a positive definite operator $\eta=e^{-\theta p}$ and the Hamiltonian is linked with that of a Hermitian Hamiltonian $h$ via an imaginary shift of coordinate which indicates that all the eigenvalues of $H$ are real and the eigenvalue problem (\ref{e4}) reduces to solving the Mathieu equation as shown below.

 (ii) For
$V_0> .5$ we have $\eta V(x) \eta^{-1} = V(x+i \theta) \neq
V^*(x)$, where $\eta= e^{-\theta p}, \theta= tanh^{-1}\frac{1}{
2V_0}$ which indicates that in this case the Hamiltonian $H$ is
not $\eta$ pseudo-Hermitian so the reality of the eigenvalues is
not ensured.

 (iii)  For $V_0=.5$, the potential $V(x) = 2+2 e^{2 i
x}$ had been studied in \cite{r9,r25,r10}. In ref.\cite{r25}, it was shown that the spectrum of this potential that is defined on the real line ($x$ is real) is purely continuous and fills the semi axis $[0,\infty)$. There are no band gaps at all. This is consistent with the results of ref.\cite{r10}. In the latter paper it was also shown that the spectrum of the Schrodinger Hamiltonian for the above mentioned potential can be discrete if one chooses the complex contour appropriately.

 Hence it is clear that $V_0=V_0^{th}=.5$ is the transition
point after which the eigenvalues of equation (\ref{e4}) becomes
complex. This ${\cal{PT}}$ threshold $V_0^{th}$ has been
determined earlier in ref. \cite{r1,r2} using numerical
techniques.

 After making the change of variable $z=x-\frac{i}{2} tanh^{-1}\frac{1}{2V_0}$ in case of $V_0> .5$,
the equation (\ref{e4}) transforms into
\begin{equation}
\frac{d^2\psi(z;\beta,V_0)}{dz^2}+\left(2-\beta+2i\sqrt{4V_0^2-1}~
sin 2z\right)\psi(z;\beta,V_0)=0.\label{e5}
\end{equation}
Under simple coordinate translation $z \rightarrow
\frac{\pi}{4}-z$ the  equation (\ref{e5}) reduces to Mathieu
equation $ \frac{d^2\psi}{dz^2} + (a-2 i qcos2z)\psi = 0$
\cite{r13}, with characteristic value $a = 2-\beta$ and $q =
\sqrt{4V_0^2 - 1}$. A variety of numerical methods have been
employed to solve Mathieu equation with an imaginary
characteristic parameter as is the case in (\ref{e5}). Tabular
values from an old paper of Mulholland and Goldstein \cite{r14}
and Bouwkamp \cite{r15} reflect that the characteristic values $q$
would be both real and conjugate complex: the transition occurring
in the neighborhood of $q= q_0= 1.46876852$ for  solution having a
period $\pi$. The same critical(exceptional) point was obtained in
ref.\cite{r17}, while considering the large $N$ limit of the quasi
exactly solvable ${\cal{PT}}$ symmetric potential $V(x) = -(i \xi
sin 2x +N)^2$. So from equation (\ref{e5}) it follows that there
is a second critical point in the neighborhood of $V_0=V_0^{c}
\sim .888437$ after which all the eigenvalues of the equation
(\ref{e5}) becomes complex and occur in pairs as complex
conjugates. Consequently no part of the band structure remain
purely real. In the Figure 3 we have plotted the band structure
for the value of $V_0=.8885$ using Floquet analysis \cite{r16}.

Now to determine the eigenvalues of the equation (\ref{e5}) and
correspondingly the band structure of the potential given in
(\ref{e1}), we proceed as follows \cite{r23}. As the coefficient
of the eq.(\ref{e5}) are $\pi$ periodic, so according to
Floquet-Bloch theorem \cite{r16} the eigenfunctions are of the
form $\psi(z;\beta, V_0)=\phi_\nu(z;\beta, V_0) e^{i\nu z}$, where
$\phi_\nu(z+\pi)=\phi_\nu(z)$ and $\nu $ stands for the real Bloch
momentum. Since $\phi_\nu(z)$ is periodic so the entire solution
of equation (\ref{e5}) can be expanded in the Fourier series
\begin{equation}
\psi(z,\beta) =  \sum_{k\in \mathbb{Z}} c_{2k} (\nu;\beta,V_0)
e^{i(\nu+2k) z}\label{e6}
\end{equation}
which, inserted into eq.(\ref{e5}), gives the recurrence relation
\begin{equation}
 c_{2k}+\zeta_{2k}[c_{2(k-1)}-c_{2(k+1)}]
=0,~~\zeta_{2k}=\frac{-\sqrt{4V_0^2-1}}{[(\nu+2k)^2+\beta-2]},~~\forall
k\in \mathbb{Z}\label{e7}
\end{equation}
for the coefficients $c_{2k}(\beta,\nu)$. For any
finite(truncated) upper limit $n\in \mathbb{Z}$, eq. (\ref{e7})
can be written as a matrix equation
\begin{equation}
\mathcal{M}_n(\nu;\beta,V_0) \mathcal{C}_n^t = 0\label{e8}
\end{equation}
where $\mathcal{C}_n=(c_{-2n},...,c_{-2},c_0,c_2,...,c_{2n})$ and
$\mathcal{M}_n(\nu;\beta,V_0)$ is a $2n+1\times 2n+1$ matrix given
by
\begin{equation}
\mathcal{M}_n(\nu;\beta,V_0) =\left(%
\begin{array}{ccccccccc}
  1 & -\zeta_{2n} &  &  &  &  &  & &  \\
  \zeta_{2n-2} & . & . & . & . & . & . & . &  \\
   & . & \zeta_2 & 1 & -\zeta_2 & 0 &0 & . &  \\
   & . & 0 & \zeta_0 & 1 & -\zeta_0 & 0 & . &  \\
   & . & 0 & 0 & \zeta{-2} & 1 & -\zeta_{-2} & . &  \\
   & . & . & . & . & . & . & . & -\zeta_{-2n+2} \\
   &  &  &  &  &  &  & \zeta_{-2n} & 1 \\
\end{array}%
\right).\label{e9}
\end{equation}
For the finite dimensional case, finding the non-trivial solutions
of the homogeneous system of eqn.(\ref{e8}) is equivalent to solve
\begin{equation}
det(\mathcal{M}_n) = 0.\label{e10}
\end{equation}
The solution of the above eqn.(\ref{e10}) gives the relation
between the energy eigenvalues $\beta$ and bloch momentum $\nu$.
Figure 2 and Figure 3 have been drawn taking $n=3$.

\begin{figure}
 \epsfxsize=2.5 in \epsfysize=1.75 in
 {\centerline{\epsfbox{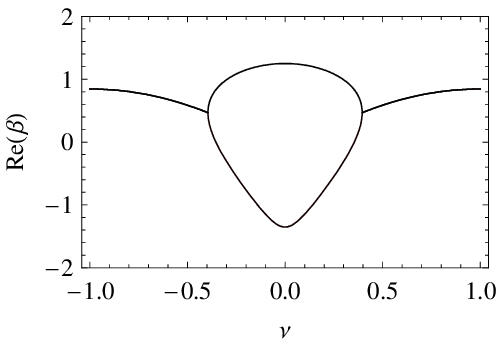} \epsfxsize=2.5 in \epsfysize=1.75 in \epsfbox{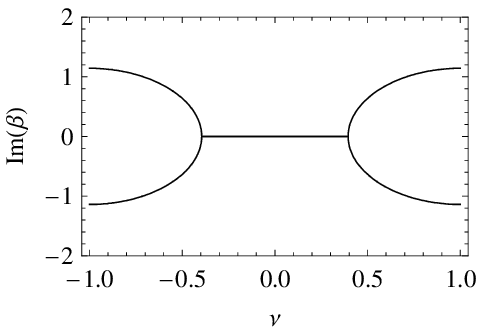}}}
 \caption{Band structure for the $\mathcal{PT}$ invariant potential $V(x)=4 cos^2 x + 4iV_0 sin 2x$ for $V_0$=.75}
 \end{figure}

 \begin{figure}
{\epsfxsize=2.55 in \epsfysize=1.5 in
{\centerline{\epsfbox{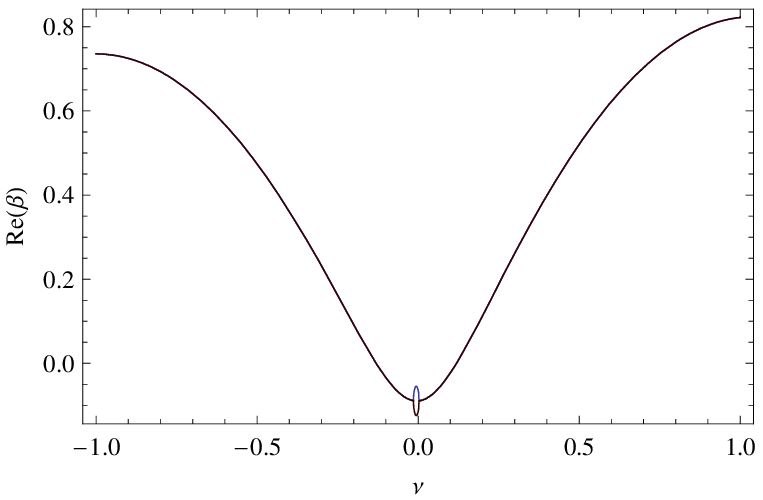} \epsfxsize=2.5 in \epsfysize=1.5 in
\epsfbox{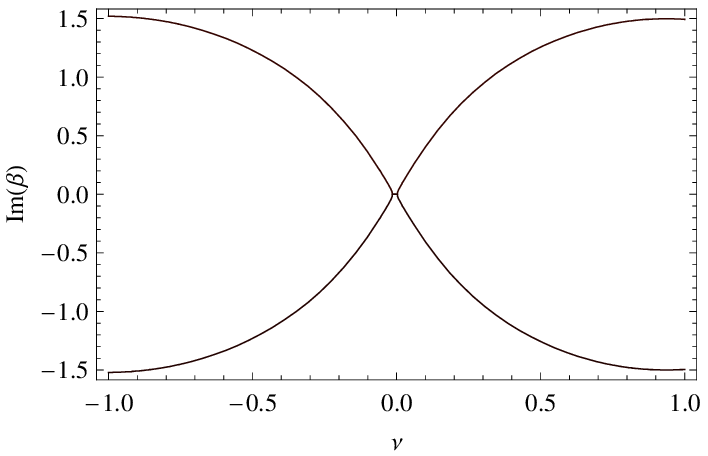}}}}\caption{Band structure for the $\mathcal{PT}$
invariant potential $V(x)=4 cos^2 x + 4iV_0 sin 2x$ for
$V_0$=.8885}
 \end{figure}

\newpage
In conclusion,

 (i) we have shown that for a positive definite operator $\eta$ the potential (\ref{e1}) is
$\eta$ pseudo-Hermitian for $V_0<.5$ which ensures that all the
band edges are real. It has been also that for $V_0<.5$ the eigenvalue problem (\ref{e4}) reduces to solving the Mathieu equation. With the aid of $\eta$ pseudo-Hermiticity,
the $\cal{PT}$ threshold  for the potential under consideration,
is obtained as $V_0^{th}=.5$. This was previously obtained in
ref.\cite{r1,r2} using numerical technique.

 (ii) The existence of a second critical point $V_0^c\sim .888437$ is established.
Beyond this critical value no part of the band structure remains
real. Numerical calculations also support this as is shown in
Figure 3. At this point, it is important to note that in
ref.\cite{r2}, the authors have shown that it is possible to find
stationary self trapped modes (with real eigenvalues)
of the nonlinear equation $\frac{d^2\psi}{dx^2}+V(x)\psi +
|\psi|^2 \psi=\beta\psi$ above the ${\cal{PT}}$ threshold
$V_0^{th}=.5.$ This is because of the fact that the band structure
still remains real for some values of bloch momentum $\nu$ even
above the ${\cal{PT}}$ threshold. Stability analysis however
revealed that the above class of lattice solitons are unstable.
The possible physical implication of the second critical point
$V_0^c$ may be that there would not exist any soliton (neither
stable nor unstable) solutions of the above nonlinear equation
beyond $V_0^c$. It will be interesting to study this physical
aspect and other related physical properties of the optical
lattice beyond this second critical point $V_0^c.$

\end{document}